\documentclass[10pt, journal, onecolumn]{IEEEtran}
\usepackage{graphicx}
\IEEEoverridecommandlockouts
\usepackage{amsmath,stfloats}
\include{psfig}
\usepackage{caption}
\usepackage{epsfig}
\usepackage{enumerate}
\usepackage{multicol}
\usepackage{array}
\usepackage{psfrag}
\usepackage{amssymb}
\usepackage{mathdots}
\include{psfig}
\usepackage{color}
\usepackage{pstricks,pst-node,pst-text,pst-3d,pst-plot}
\usepackage{psfrag}

\begin{document}

\title{The asymmetric DoF Region for the 3-user $M\times N$ Interference Channel}


\author{\IEEEauthorblockN{Mohamed Khalil\IEEEauthorrefmark{1}, Tamer Khattab\IEEEauthorrefmark{1}, Amr El-Keyi\IEEEauthorrefmark{2}, Mohammed
Nafie\IEEEauthorrefmark{2}}
    \\\IEEEauthorblockA{\IEEEauthorrefmark{1}Qatar University
    \\\{mohamed.amir, Tkhattab\}@qu.edu.eg,}
   \\ \IEEEauthorblockA{\IEEEauthorrefmark{2}Wireless Intelligent Networks Center (WINC) \\
Nile University, Cairo, Egypt.
    \\\{aelkeyi, mnafie\}@nileuniversity.edu.eg}

\thanks{This work was supported in part by a grant from Qatar national research fund . Mohammed Nafie is also affiliated with Faculty
of Engineering, Cairo University.}}

\maketitle

\vspace{-1mm}
\begin{abstract}
In this paper, the 3-user Gaussian MIMO interference channel is considered. The asymmetric distribution of the DoF,  where different users have different number of DoF, is studied. Two cases are presented, the first is when all transmitters and receivers have equal number of antennas $M$, the other when the transmitters have $M$ antennas each, while the receivers have $N$ antennas each. It is assumed that the
channel coefficients are constant and known to all transmitters
and receivers.  The region of the achievable DoF tuple
$(d_1; d_2; d_3)$ is presented.
\end{abstract}
\IEEEpeerreviewmaketitle

\section{Summary}

%


The DoF region of 3-user $M\times N$ Gaussian MIMO interference channel was presented in \cite{ours},\cite{chains}, where \cite{chains} derived the upperbound and showed that the scheme presented in \cite{ours} is DoF optimal. All previous work assumed that all transmitters
have the same number of DoF. The need to identify the asymmetric DoF region 
comes from the reality the different nodes require different rates and
have different channel strength at different times.\\
We consider the 3 user $M\times N$ interference channel as defined in the system model of \cite{ours}.
We assume full CSI knowledge at all transmitters. We use the notation $<A>_b$ to for the modulo of $A$ to the base $b$.\\

First we consider the case when $M=N$. Let $d_i$ be the DoF of user $i$ and $I_i$ be the interference at the $i$th receiver, we assume without loss of generality that 
First we consider the case when $M=N$. Let $d_i$ be the DoF of user $i$ and $I_i$ be the interference at the $i$th receiver, we assume without loss of generality that 
\begin{equation}
d_1 \geq d_2 \geq d_3  
\end{equation}
\\
For user 1 to be able decode its message, the signal space and interference space must be distinct at receiver 1 or,
\begin{eqnarray}
d_1+ I_1 \leq M\label{equalno1}\\
I_1=\text{max}(d_2, d_3)\\
d_1+ \text{max}(d_2+ d_3)  \leq M\label{equalno2}
\end{eqnarray}
\\
The interference space size is dominated by $d_1$ at receivers two and three because $d_1d_1 \geq \text{max}(d_2,d_3) $
then,
\begin{eqnarray}
I_2=I_3=d_1\\
\end{eqnarray}
Then the DoF region of the asymmetric case is,
\begin{eqnarray}
d_i+ d_j \leq M\label{equalno2},  i=1,2,3, j=1,2,3, i\neq j 
\end{eqnarray}
\\
From (\ref{equalno1},\ref{equalno2}), it is easy to see that an increase of $d^+$ to $d_i$ would result in a decrease in the DoF of other users, each by $d^+$ or a total loss in the sum DoF by $d^+$.\\

Next. we consider the general case $M\neq N$. We assume-without loss of generality-that M is larger than
N. 
In the proposed schemes, the nullspaces of the channels to
the non intended receivers are used in combination with interference alignment to mitigate the effect of
interference.
In \cite{ours}, the optimal achievable scheme for the symmetric DoF of network was presented, we will use the scheme used there.
The scheme aligns $(d_1,d_2,d_3)$ DoF of transmitters (1,2,3), respectively at $(I_1 , I_2, I_3)$ Interference dimensions at receivers (1,2,3) respectively.
The scheme in \cite{ours} aligns only portions of the signal in form
of alignment tree, this tree starts at one transmitter and ends at one receiver's nullspace and in between interference alignment is carried out at different receivers. The tree has length $L$ which may be larger than three or in other words, the tree may pass through any transmitter or any receiver multiple times, and the number of DoF achieved by any transmitter is the number of times the tree passed through it multiplied by the number of DoF in one tree branch ($d_o$).\\

As in \cite{ours}, $d_o$ is governed by the following equation
\begin {equation}
d_o=((L+1) M-(L+2)N)^+ 
\end{equation}
Each transmitter can be the root of only one tree, so we have maximum of 3 trees only.
Fig. (1) shows an example for  a tree with $L=1$, and $\boldsymbol{V}_i=[\boldsymbol{V}_i^1 \boldsymbol{V}_i^2 \boldsymbol{V}_i^3]$.\\

%

\begin{figure}[]
  \begin{center}\label{tree}
 \includegraphics[width=.50\textwidth]{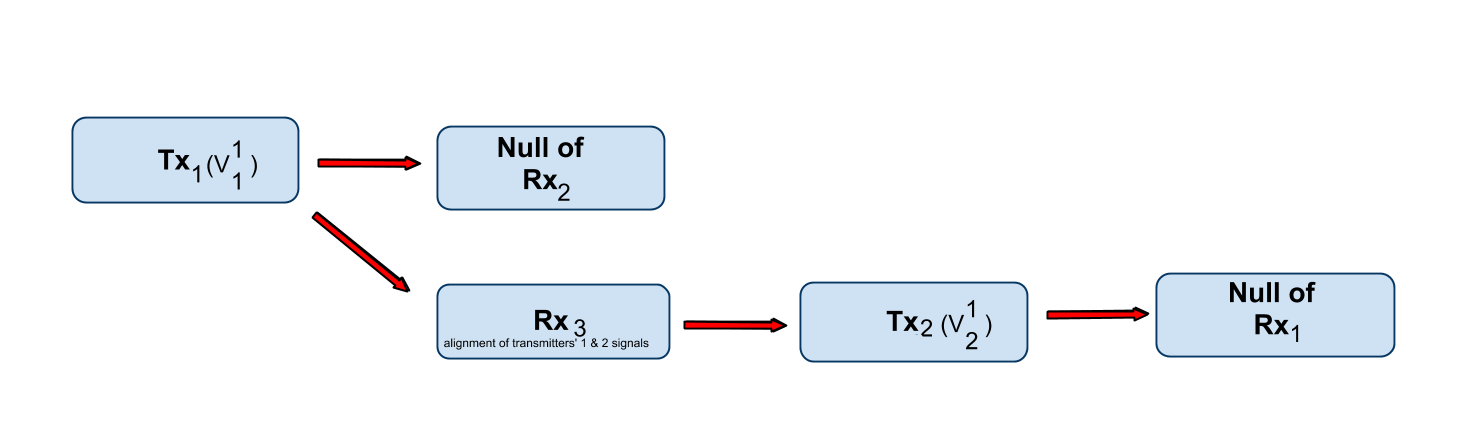}
 \caption{Alignment tree for $L=1$.} \hspace{-10cm}
  \end{center} 
	\vspace{-6mm}
  \end{figure}

The scheme's achievable DoF is governed by the two constraints, 
\begin{enumerate}
\item receiver space constraint 
\begin{equation}\label{firstconst}
d_i+I_i \leq N
\end{equation}
\item Nullspace and alignment constraint 
\begin{equation}\label{secconst}
d_o \leq ((L+1) M-(L+2)N)^+ 
\end{equation}
\end{enumerate}

In the following analysis, we will explain the schemes assuming that $d_o=1$ and the scheme would be valid for any $d_o \in Z^+$ with DoF and interference multiplied by the factor $d_o$. We will consider the sets $(d_1,d_2,d_3)$ and $(I_1 , I_2, I_3)$ in the case of $d_o=1$ as basis sets for the DoF and interference achieved at $d_o \in Z^+$.\\

The achievability scheme used depends on the ratio $\frac{M}{N}$, the region is divided into two main regions.\\

A. Lossless DoF redistribution region

We provide an achievable scheme that does not result in the decrease in the sum DoF.
\begin{enumerate}[i.]

\item \textit{$\frac{M}{N}\in[\frac{2L+3}{2L+1}, \frac{L+1}{L}[$} \hspace{4mm} $\forall L=1,4,7,...$ \\
\vspace{2mm}

The redistribution for this region is governed by the following equations.

\begin {eqnarray}
d_1=  \bigg(\frac{(L+2)}{3}a_1+ \frac{(L-1)}{3}a_2+ \frac{(L+2)}{3}a_3\bigg)\\
d_2=  \bigg(\frac{(L+2)}{3}a_1+ \frac{(L+2)}{3}a_2+ \frac{(L-1)}{3}a_3\bigg)\\
d_3=  \bigg(\frac{(L-1)}{3}a_1+ \frac{(L+2)}{3}a_2+ \frac{(L+2)}{3}a_3\bigg)\\
\end{eqnarray}

where the redistribution factor $a_i$ is bounded by $d_o$
\begin{equation}
a_i \leq d_o \forall \hspace{3mm} i=1,2,3.
\end{equation}

and the sum DoF is bounded by the sum DoF of the symmetric DoF case.

\begin{equation}
\sum_i d_i \leq 3(L+1)/(L+2)
\end{equation} 

The increase in one transmitter's DoF is governed by two factors, the number of available dimensions for transmission by the scheme after achieving the nullspace constraint (\ref{secconst})
and maximum DoF to interference space $\delta$ that can be generated by a given tree $l$.
\begin{equation}\label{dd}
\delta=\max_{l,i} (\frac{d_i^l}{I_i^l})
\end{equation}

with,

\begin{equation}
d_\text{max}+I_\text{min} \leq N
\end{equation}

\begin{figure}\label{LL}
  \begin{center}
 \includegraphics[width=.50\textwidth]{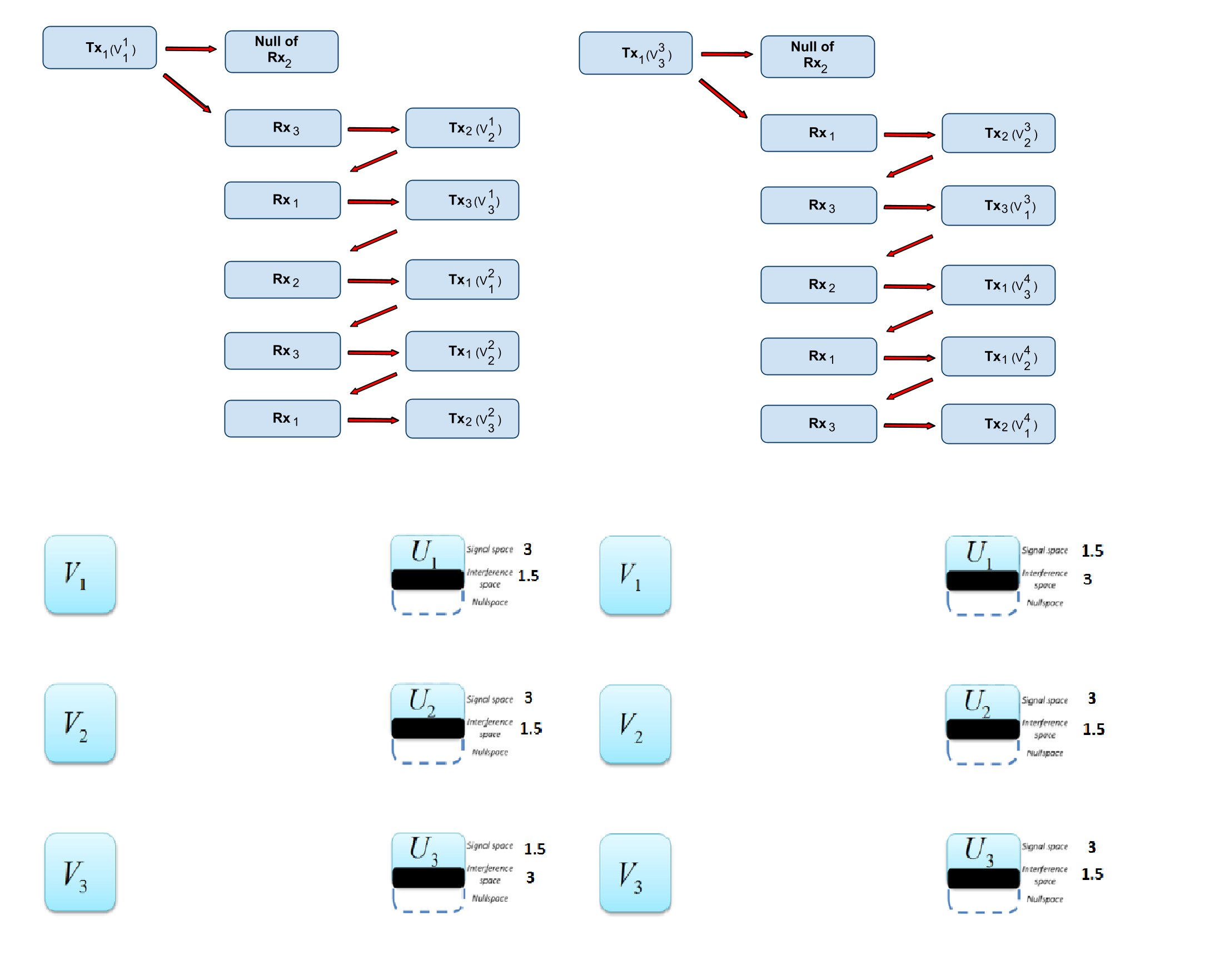}\vspace{-2mm}
 \caption{Alignment trees for L=4}
  \label{fig IA}
  \end{center} \vspace{-6mm}
  \end{figure}

An example for the scheme with $M=11.25$ and $N=9$ is explained next. The normalized fractional value of $M$ and $N$ is considered only as a basis set for all integer real values that has $\frac{M}{N}=\frac{11.25}{9}$. The same scheme applies with others values multiplied by $\frac{M}{11.25}$ or $\frac{M}{9}$. The distribution of DoF and interference for a tree with $L=4$ is $(2,2,1)$ DoF and $(2,1,1)$ interference dimensions. For this network the minimum achievable 
interference $(I_\text{min})$ is $1$ and the maximum achievable DoF $(d_\text{max})$ is $2$, the maximum DoF to interference ratio achieved for any transmitter is 
equal to 2.
If we used trees with the later ratio, and to achieve the first constraint \ref{firstconst} we have 
\begin{equation}
d \leq d_\text{max}\frac{N}{(d_\text{max}+I_\text{min})}
\end{equation}
 for $<L>_3=1$ ,$d_\text{max} =2$, $I_\text{min}=1$ and 

\begin{equation}
d \leq \frac{2N}{3}
\end{equation}
 Given that $N=9$, so the max achievable DoF for any transmitter is $\frac{2\times9}{3}=6$. 

For $M=11.25$ and $N=9$, from equation (\ref{secconst}) we have $d_o \leq 2.25$, this means that the available space that achieve the alignment constraint is (4.5, 4.5, 2.25) for the first tree and (2.25, 4.5, 4.5) for the second one. In fig.(2) we show DoF assignment used to achieve the DoF of the network, it is easy to see that the sum DoF for the set (4.5, 6, 4.5) DoF in fig.(2) is equal to the sum DoF of the symmetric network or 15.\\

\item \textit{$\frac{M}{N}\in[\frac{L+2}{L+1}, \frac{2L+3}{2L+1}[$} \hspace{4mm} $\forall L=1,4,7,...$ \\
\vspace{2mm}

The redistribution scheme depends on using trees with $L_1=L-1$ and $L_2=L$, the increase of one user DoF depends on the fraction of DoF assigned to $L_1$. Also, the number of DoF lost depends on the fraction assigned to trees of $L_1$, because these trees has a lesser DoF-to-interference ratio.\\

\end{enumerate}
B. Lossy DoF redistribution region

For this region the achievable scheme results in a decrease in the sum DoF after redistribution.
\begin{enumerate}[i.]

\item \textit{$\frac{M}{N}\in[\frac{2L+3}{2L+1}, \frac{L+1}{L}[$} \hspace{4mm} $\forall L=2,5,8,...$ \\
\vspace{3mm}
For this case any DoF redistribution would be lossy, because the signal basis set is equal to (1,1,1), then all trees would 
 give the same DoF distribution. 
\\

\item \textit{$\frac{M}{N}\in[\frac{2L+3}{2L+1}, \frac{L+1}{L}[$} \hspace{4mm} $\forall  L=3,6,9...$ 
\vspace{3mm}
The redistribution for this region is lossy as well, because the interference basis set is equal to (1,1,1), then all tree would 
cause the same interference pattern. Any DoF redistribution between the scheme trees would not change the interference pattern at the receivers. As result, the empty space at all receivers for signal decoding would not change. 

\end{enumerate}


\end{document}